\documentstyle[12pt,aps,epsfig]{revtex}

\begin{document}

\title{
Scattering of electrons by impurities within the framework of two band
model of order parameter anisotropy. }
\author{ P.I. Arseev, N.K. Fedorov, S.O. Loiko.}
\address{P.N. Lebedev Physical Institute of RAS, Moscow 119991, Russia.}
\maketitle

\begin{abstract}
The impurity concentration dependence of superconducting
transition temperature $T_c$ is studied within the framework of
two band model of order parameter anisotropy. Both intraband and
interband electron scattering by nonmagnetic and magnetic
impurities are taken into account. It is demonstrated that at
various values of model parameters both types of impurity
concentration dependence of $T_c$ are possible: weak dependence
typical of the model with $s$-wave order parameter and strong
suppression of $T_c$ by impurities with critical impurity
concentration at which $T_c=0$. It is found that in some cases
there is a possibility of a rise of critical temperature with
increasing impurity concentration. The obtained results are
consistent with existing experimental data for the
high-temperature superconductors.
\end{abstract}

The most important and still unresolved problem in the theory of
high-temperature superconductivity (HTSC) is the mechanism of
formation of electron-electron attraction responsible for unusual
properties of these compounds. In this connection an investigation
of order parameter symmetry is of great importance. On the one
hand, a variety of experimental data suggests strong anisotropy
of order parameter \cite{ding,zhao}. On the other hand, a number
of experiments displays the behavior typical of models with
$s$-wave order parameter \cite{pon,suzuk}. It was shown earlier
that order parameter anisotropy can be explained by symmetry
properties of superconductor crystal lattice within the framework
of the universal multi-band model \cite{anis}. The approach offers
a non-contradictory explanation of a number of experimental data
for high-temperature superconductors irrespective of the nature
of superconducting pairing.

Order parameter symmetry proves to be essential if the effect of
impurities on critical temperature is investigated. Single-band
models (BCS-like) with $d$-wave order parameter display strong
reduction of $T_c$ with increasing nonmagnetic impurity
concentration \cite{gor-kal,sigr,maki}. Analogous effect on $T_c$
is produced by electron scattering by magnetic impurities in model
with $s$-wave order parameter \cite{abr-gor,open}. In either case
critical temperature vanish at certain impurity concentration.
However, the suppression of superconductivity induced by
scattering from nonmagnetic impurities is inconsistent with
existing experimental data \cite{radtke}, including the earlier
ones, when samples were not pure enough. There are several
different ways of possible explanation of weak reduction of
critical temperature with increasing nonmagnetic impurity
concentration in models with strongly anisotropic order parameter.
The inclusion of anisotropy of scattering matrix element in the
models with anisotropic order parameter \cite{dolg,nagi,haran} can
lead to weak impurity concentration dependence of $T_c$. Relative
arrangement of nodes of order parameter and those of impurity
potential plays an important role here. It was demonstrated in
\cite{gork}, that non-trivial topology of Fermi surface can
strongly affect superconducting properties, and in particular
inhibit reduction of $T_c$ with increasing impurity concentration.
There is a separate class of models describing formation of order
parameter anisotropy, namely, multi-band models.
Although scattering by impurities was repeatedly investigated within the
framework of these models \cite{dolg,gork,mosk,kusak,keller,comb,mazin}, a
number of questions concerning consideration of real crystal structure and
symmetry properties of electron spectrum remains unsolved.

The present article is concerned with effect of electron
scattering by impurities on critical temperature within the
framework of previously proposed two-band model of an order
parameter anisotropy \cite{anis}. The width of one band is
assumed to be much greater than that of another band. Hamiltonian
of the model has the form

\begin{eqnarray}
H_0 &  =  & \sum_{i,j,\alpha }(\varepsilon_{a0}\delta_{ij} +
t_{ij}^a)a_{i,\alpha }^{+}a^{}_{j,\alpha } \nonumber \\
& + & \sum_{i,j,\alpha }(\varepsilon_{c0}\delta_{ij}+t_{ij}^c)c_{i,\alpha
}^{+}c^{}_{j,\alpha }
             \nonumber             \\
 & +  & \sum_{i,j,\alpha }(W_{ij}a^{+}_{i,\alpha}
c^{}_{j,\alpha}+h.c.)
                              \nonumber      \\
     &  +  &
    U_{a}\sum_{i}a_{i\downarrow}^{+}a_{i\downarrow}
 a_{i\uparrow}^{+}a_{i\uparrow}
\nonumber      \\
     &  +  & U_c\sum_{i}c_{i\downarrow}^{+}c_{i\downarrow}c_{i\uparrow}^{+}
      c_{i\uparrow}  \quad ,
\end{eqnarray}
 where the operators $a_{i,\alpha }^{+}$ and $c_{i,\alpha }^{+}$
 create electrons on the $i$-th site with spin $\alpha$ in the wide and
 the narrow bands respectively;  $\varepsilon_{a0}$ and $\varepsilon_{c0}$
 are the energies of levels forming the wide and the narrow bands; $t_{ij}^a$ and $t_{ij}^c$
 are the matrix elements of single-particle transitions between
 sites in the wide and the narrow bands; $W_{ij}$ is the matrix element of
 single-particle interband transitions. It is significant that
 since $W_{ij}$ depend on the indices of different sites, in the $\bf
k$- representation the parameter $W$ must be a function of $\bf
k$. It is assumed that superconductivity is caused by isotropic
attraction in the narrow band $U_c<0$ and there is isotropic
effective interaction (either repulsive or attractive) between
electrons in the wide band $U_a$.

The interaction of electrons with impurities is described by the
following Hamiltonian

\begin{eqnarray}
\hat{H}_{imp} &  = & \sum_{{\bf k},{\bf k'},\alpha,\beta}U_{aa}({\bf
k},\alpha;{\bf k}',\beta)a^+_{{\bf k},\alpha}a^{}_{{\bf k}',\beta}
\nonumber      \\
     &  +  &
\sum_{{\bf k},{\bf k}',\alpha,\beta}U_{cc}({\bf k},\alpha;{\bf
k}',\beta)c^+_{{\bf
k},\alpha}c^{}_{{\bf k}',\beta} \nonumber \\
&  +  &  \sum_{{\bf k},{\bf k}',\alpha,\beta}(U_{ac}({\bf k},\alpha;{\bf
k}',\beta)a^+_{{\bf k},\alpha}c^{}_{{\bf k}',\beta} \nonumber \\
&    & \qquad \qquad\qquad\qquad\qquad+ h.c.),
\end{eqnarray}
where $$U_{ij}({\bf k},\alpha;{\bf k}',\beta)=\sum_{l} u_{ij}({\bf
k},\alpha;{\bf k}',\beta)e^{-i({\bf k}-{\bf k}'){\bf R}_l}$$ is the matrix
element for electron scattering by impurities from the state $({\bf
k}',\beta)$ in $j$ band to the state $({\bf k},\alpha)$ in $i$ band
$(i,j=a,c)$, ${\bf R}_l$ is the position of $l$-th impurity . In general,
$$u_{ij}({\bf k},\alpha;{\bf k}',\beta)=(u^n_{ij}({\bf k},{\bf
k}')+ u^m_{ij}({\bf k},{\bf k}'))\delta_{ \alpha \beta} $$
$$\qquad+\displaystyle{\frac{1}{2}}J_{ij}({\bf k},{\bf k}'){\bf S}_l{\bf
\sigma}_{\alpha \beta}$$
Here ${\bf S}_l$ is the spin of magnetic impurity
located at ${\bf R}_l$, ${\bf \sigma}_{\alpha \beta}$ are the matrix
elements of Pauli matrices,
$$u^{n(m)}_{ij}({\bf k},{\bf k}')=\int \psi_{{\bf k}}^{(i)*}({\bf
r})u^{n(m)}({\bf r})\psi_{{\bf k}'}^{(j)}({\bf r})d^3{\bf r}\,,$$
$$J_{ij}({\bf k},{\bf k}')=\int \psi_{{\bf k}}^{(i)*}({\bf
r})J({\bf r})\psi_{{\bf k}'}^{(j)}({\bf r})d^3{\bf r}\,,$$ where
$u^{n(m)}({\bf r})$ is the potential of nonmagnetic (magnetic)
impurity, $J({\bf r})$ is the exchange interaction of electron
with magnetic impurity, $\psi_{{\bf k}}^{(i)}$ is Bloch function
of electron in $i$ band.

The impurity atoms will be assumed to be randomly distributed over
the crystal lattice and the mean distance between them is much
greater in comparison with interatomic distances. Let us consider
short-range impurity potentials, so that the matrix elements
$u^{n(m)}_{ij}({\bf k} ,{\bf k}')$ and $J_{ij}({\bf k},{\bf k}')$
do not depend on quasimomentum and are equal to $u^{n(m)}_{ij}$
and $J_{ij}$ respectively. Thus, both intraband and interband
electron scattering by impurities are isotropic. All calculations
will be performed within Born approximation.

Let us introduce the following Matsubara Green's functions:

$$
G^a_{i,j}(\tau,\tau')=-i\left\langle T\ a_{i,\alpha
}(\tau)a^{+}_{j,\alpha  }(\tau')\right\rangle,
    $$
    $$
G^c_{i,j}(\tau,\tau')=-i\left\langle T\ c_{i,\alpha }(\tau)c^{+}_{j,\alpha
}(\tau')\right\rangle,    $$
$$ D_{i,j}(\tau,\tau')=-i\left\langle T\
a_{i,\alpha}(\tau)c^{+}_{j,\alpha }(\tau')\right\rangle, $$
$$ i \sigma^y_{\alpha \beta} F^{a+}_{i,j}(\tau,\tau')=-i\left\langle T\
a^{+}_{i,\alpha}(\tau)a^{+}_{j,\beta}(\tau')\right\rangle,  $$
$$
i \sigma^y_{\alpha \beta} F^{c+}_{i,j}(\tau,\tau')=-i\left\langle T\
c^{+}_{i,\alpha}(\tau)c^{+}_{j,\beta}(\tau')\right\rangle,  $$
$$
i \sigma^y_{\alpha \beta}B^{+}_{i,j}(\tau,\tau')=-i\left\langle T\
a^{+}_{i,\alpha}(\tau)c^{+}_{j,\beta}(\tau')\right\rangle.
$$

The system of equations for these functions can be obtained by
using equations of motion for the operators $a_{i,\alpha}$ and
$c_{i,\alpha}$. Then  make transform to $({\bf
k},\omega)$-representation and perform Abrikosov-Gorkov procedure
of averaging over impurity positions and, if impurities are
magnetic, over their spin orientations \cite{abr-gor}. As a
result the system of equations for the narrow band Green's
functions takes the following form:
\begin{eqnarray}
\left( i\omega_c - \varepsilon_{c}(\bf k)\right) G_{c}({\bf k},\omega)
-\widetilde{W}^{*}_{ca}({\bf k})D({\bf k},\omega) \qquad \nonumber \\
\qquad + \Delta _{c}F_{c}^{+}({\bf k},\omega)+\Delta_{ca} B^+(-{\bf
k},-\omega)=1,
\end{eqnarray}
\begin{eqnarray}
\left( i\omega_a - \varepsilon_{a}({\bf k})\right) D({\bf k},\omega)
-\widetilde{W}^{*}_{ac}({\bf k})G_{c}({\bf k},\omega) \qquad \nonumber \\
\qquad +  \Delta _aB^{+}(-{\bf k},-\omega)+\Delta_{ac} F_{c}^+({\bf
k},\omega)=0,
\end{eqnarray}
\begin{eqnarray}
\left(i \omega_c^* + \varepsilon_{c}(\bf k)\right) F_{c}^{+}({\bf
k},\omega) +\widetilde{W}_{ca}({\bf k})B^{+}(-{\bf k},-\omega)  \nonumber \\
+ \Delta_{c}^{+}G_{c}({\bf k},\omega)+\Delta_{ca}^+ D({\bf k},\omega)=0,
\qquad \quad
\end{eqnarray}
\begin{eqnarray}
\left(i \omega_a^* + \varepsilon_{a}({\bf k})\right) B^{+}(-{\bf
k},-\omega) +\widetilde{W}_{ac}({\bf k})F_{c}^{+}({\bf k},\omega)  \nonumber \\
 + \Delta_a^{+}D({\bf k},\omega)+\Delta_{ac}^+ G_c({\bf
k},\omega)=0. \qquad \quad
\end{eqnarray}
The system of equations for the wide band Green's functions can be
obtained from (3-6) by interchanging of band indices $a$ and $c$.

Here new parameters  are introduced, among which are renormalized
frequencies $\omega_a$, $\omega_c$ and anomalous means $\Delta_a$,
$\Delta_c$ appearing also in single-band models:

\begin{eqnarray}
i \omega_c = i \omega &  -  & \frac{n^{(n)}
|u_{cc}^{n}|^2+n^{(m)}(|u_{cc}^{m}|^2+|v_{cc}|^2)}{(2 \pi)^3} \nonumber \\
& \times&  \int d^3{\bf k}\,G_c({\bf k},\omega)
\nonumber \\
&  -  & \frac{n^{(n)}|u^n_{ac}|^2+n^{(m)}(|u^m_{ac}|^2+|v_{ac}|^2)}{(2
\pi)^3} \nonumber \\
&\times&  \int d^3{\bf k}\,G_a({\bf k},\omega),
\end{eqnarray}
\begin{eqnarray}
\Delta_c= \Delta_{c0} & + & \frac{n^{(n)}
|u_{cc}^{n}|^2+n^{(m)}(|u_{cc}^{m}|^2-|v_{cc}|^2)}{(2 \pi)^3} \nonumber \\
& \times &  \int d^3{\bf k}\,F_c({\bf k},\omega) \nonumber \\ & + &
\frac{n^{(n)}|u^n_{ac}|^2+n^{(m)}(|u^m_{ac}|^{2}-|v_{ac}|^2)}{(2
\pi)^3}\nonumber \\
& \times &  \int d^3{\bf k}\,F_a({\bf k},\omega),
\end{eqnarray}
where $\Delta_{c0}=-U_c \langle c_{i\uparrow}c_{i\downarrow}
\rangle $, $n^{(n)}$ and $n^{(m)}$ are concentrations of
nonmagnetic and magnetic impurities respectively,
$|v_{ij}|^2=\displaystyle{\frac14} J_{ij}^2 S(S+1)$ $(i,j=a,c)$,
$\omega=(2n+1) \pi T$.

In addition, electron scattering by impurities leads to renormalization of
matrix element of hybridization $W$, as well as to appearance of interband
anomalous self-energy $\Delta_{ca}$:
$$
\widetilde{W}^*_{ca}=W^* +
\frac{n^{(n)}u^n_{cc}u^n_{ca}+n^{(m)}(u^m_{cc}u^m_{ca}+v_{cc}v_{ca})}{(2
\pi)^3} $$
$$
 \times  \int d^3{\bf k}\,G_c({\bf k},\omega) \qquad
 $$
$$
\qquad \qquad \quad+
\frac{n^{(n)}u^n_{ca}u^n_{aa}+n^{(m)}(u^m_{ca}u^m_{aa}+v_{ca}v_{aa})}{(2
\pi)^3}  $$
$$
 \times  \int d^3{\bf k}\,G_a({\bf k},\omega),  \qquad  \eqno(9)
$$

$$
\Delta_{ca}  =
\frac{n^{(n)}u^n_{cc}u^{n*}_{ca}+n^{(m)}(u^m_{cc}u^{m*}_{ca}-v_{cc}v^*_{ca})}{(2
\pi)^3} \qquad $$
$$
 \times  \int d^3{\bf k}\,F_c({\bf k},\omega) \qquad\qquad\qquad $$
$$  +
\frac{n^{(n)}u^n_{ca}u^{n*}_{aa}+n^{(m)}(u^m_{ca}u^{m*}_{aa}-v_{ca}v^*_{aa})}{(2
\pi)^3} $$
$$  \times  \int d^3{\bf k}\,F_a({\bf k},\omega). \qquad\qquad\qquad\eqno(10)
$$

The expressions for $\omega_a$, $\Delta_a$, $\widetilde{W}_{ac}$
and $\Delta_{ac}$ can be obtained from (7-10) by interchanging of
band indices $a$ and $c$.

The parameters $\Delta_{a0}$ and $\Delta_{c0}$ obey the
self-consistency equations
$$
\Delta_{a0}=-U_a T \sum_{n} \int \frac{d^3{\bf k}}{(2 \pi)^3}\,F_{a}({\bf
k},\omega), \eqno(11)
$$
$$
\Delta_{c0}=|U_c| T \sum_{n} \int \frac{d^3{\bf k}}{(2 \pi)^3}\,F_{c}({\bf
k},\omega). \eqno(12)
$$

If impurities are absent in a superconductor
$(n^{(n)}=n^{(m)}=0)$, the system (4-7) is reduced to the Gorkov's
system of equations for the functions $G_{c}$ and $F^{+}_{c}$ with
effective order parameter
$$\tilde{\Delta}_c({\bf k})=\frac{W^2({\bf
k})\Delta_a}{\varepsilon^2_a({\bf k})+\Delta_a^2}+\Delta_c \quad.
\eqno(13)$$

The essential feature of considered model is the strong anisotropy of the
matrix element of single-particle interband hybridization $W({\bf k})$
which has lines of nodes along the diagonals of Brillouin zone for
different types of symmetry of initial orbitals \cite{anis}. As a result
the effective order parameter (13) depends on quasimomentum in the case of
nonzero interaction in the wide band. From Eqs.(11-13) it follows that if
the interaction in the wide band is repulsive and its value is less than
the critical one, at which supeconductivity in the system is destroyed
$(U_a^{crit}>U_a>0)$, then $\Delta_a/\Delta_c<0$ and the order parameter
$\tilde{\Delta}_c({\bf k})$ changes its sign. Such a behaviour determines
the main features of the model and makes it possible to explain some
properties of HTSC compounds non-contradictory, e.g. strong anisotropy of
the order parameter and experimentally observed $s$-like behaviour of
$\displaystyle{ \frac{dI}{dV} }$\,-characteristics of $SIS$ junctions
\cite{tunn}.

The impurity concentration dependence of $T_c$ can be determined by
solving the systems of equations (3-6) and (7-12) in the limit
$\Delta_a,\Delta_c \rightarrow 0$ at $T \rightarrow T_c$.

The chemical potential will be considered to be situated in the
middle of the wide band with energy dispersion $\varepsilon_a({\bf
k})=t_a(cosk_x+cosk_y)$. The location of the narrow band with
energy dispersion $\varepsilon_{c}({\bf
k})=\varepsilon_{c0}+t_c(cosk_x+cosk_y)$ relative to $\mu$ is
defined by the parameter $\varepsilon_{c0}$. The matrix element of
hybridization, assuming different symmetries of initial orbitals,
namely of $s$- and $d$- types, has the form $W({\bf
k})=W_0(cosk_x-cosk_y)$. For numerical calculations of impurity
concentration dependencies of $T_c$ the following values of
parameters were used: $t_a=30$, $W_0=5$, $U_a=27$, $U_c=-8$
(expressed in terms of $t_{c}$).

The relative value of intraband and interband scattering matrix elements
strongly depends on location of impurity atom in a unit cell. For example,
these can be of the same order of magnitude, i.e. $u_{aa} \propto  u_{cc}
\propto u_{ac}$. However, for substitutional impurities, scattering of
electrons of one of the initial bands by impurities can overwhelm, while
interband scattering turns out to be much weaker for symmetry reasons. It
will be shown below that a behaviour of $T_c$ with increasing impurity
concentration is different for various cases.

\begin{figure}
\leavevmode \centering{
\psfig{figure=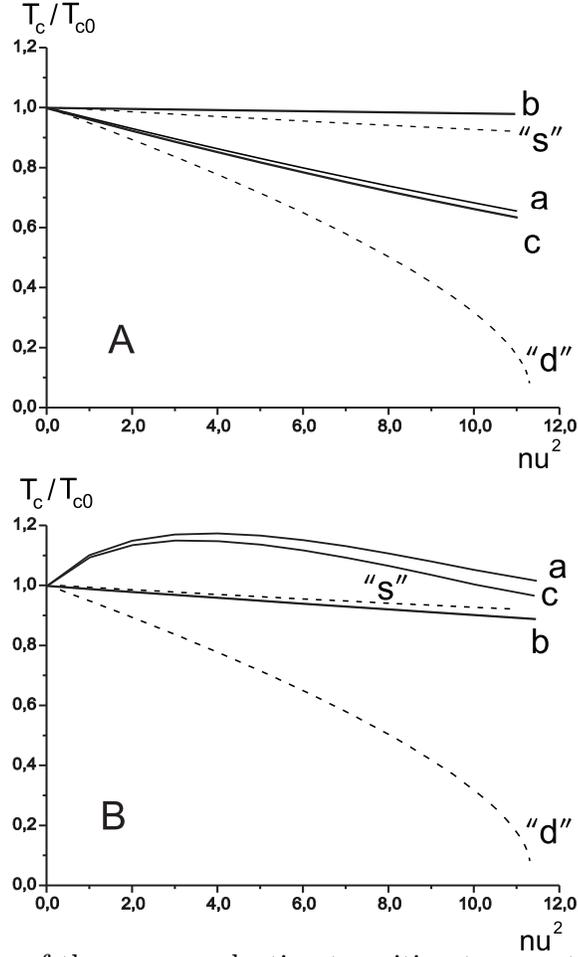}} \caption{The dependencies of the superconducting
transition temperature $T_c$ on the parameter $nu^2$ corresponding to
electron scattering by nonmagnetic impurities calculated for A)
$\varepsilon_{c0}=0$, B) $\varepsilon_{c0}=-3$ and a) $u=u_{c}$,
$u_{ac}=u_{a}=0$; b) $u=u_{a}$, $u_{ac}=0.5u_{a}$, $u_c=0$; c) $u=u_{c}$,
$u_{ac}=0.5u_{c}$, $u_{a}=0$; "s"- within the framework of single-band
model with $s$-wave order parameter; "d"- within the framework of
single-band model with $d$-wave order parameter.}
\end{figure}

In what follows the effect of electron scattering by nonmagnetic
(Fig.1) and magnetic (Fig.2) impurities on the critical
temperature depending on the location of the center of the narrow
band $\varepsilon_{c0}$ relative to the chemical potential $\mu$
is considered. The obtained $T_c(n)$ dependencies are compared
with the results calculated within the framework of single-band
models with $s$- and $d$-wave order parameters. Here mean electron
density of states of these models corresponds approximately to
that of two-band model.

Fig.1A gives the critical temperature $T_c$ as a function of
parameter $n^{(n)}u^2$ (where $u=u_{aa},u_{cc},u_{ac}$), chemical
potential being in the narrow band ($\varepsilon_{c0}=0$). As can
be seen in the figure, scattering in the narrow band has a
strongest effect on the critical temperature. Including of
interband scattering scarcely affects the result. This is
supported by the fact that interband scattering together with
intraband one in the wide band produces a weak effect on $T_c$ .
Corresponding curve coincide with the one calulated within the
framework of single-band model with $s$-wave order parameter very
closely.

If the center of the narrow band is far from the chemical potential
($\varepsilon_{c0}=-3$), the behaviour of $T_c$ with increasing impurity
concentration is changed (Fig.1B). In this case interband scattering
decreases the critical temperature while intraband scattering in the
narrow band raises it in the region of small values of parameter
$n^{(n)}u_{cc}^2$. The increase of $T_c$ at small impurity concentration
can be explained by the fact that electron density of states increases in
energy region near the chemical potential because of broadening of a peak
originating from the van Hove singularity of the initial narrow band. The
effect in this region of parameters overwhelms destruction of
superconductivity caused by scattering by impurities.

It is seen from Fig.1 that though the order parameter of
considered model changes its sign, nonmagnetic impurities
suppress superconductivity much weaker than in single-band models
with $d$-wave order parameter. There is also no critical impurity
concentration, at which the superconductivity in the system
disappears. Intraband scattering in the wide band and interband
scattering scarcely affect the critical temperature.

A series of experiments aimed at investigating  magnetic impurity
concentration dependence of $T_c$ of HTSC compounds have been
performed. As magnetic impurities $Zn$ and $Ni$ atoms
substituting for copper sites in $CuO_2$ plane were used
\cite{jayar,zhaoJPh93,tolp,kriv}. Experimental data indicates that
these impurities influence critical temperature as impurity
magnetic centers rather than by changing carrier concentration
\cite{zhaoJPh93,will}. It should be noted that while $Zn$
substitution results in strong $T_c$ suppression in the copper
oxides similar to that induced by magnetic impurities in
conventional superconductors, $Ni$ substitution was
experimentally observed to affect $T_c$ weakly \cite{zhaoJPh93}.

\begin{figure}
\centering{\epsfbox{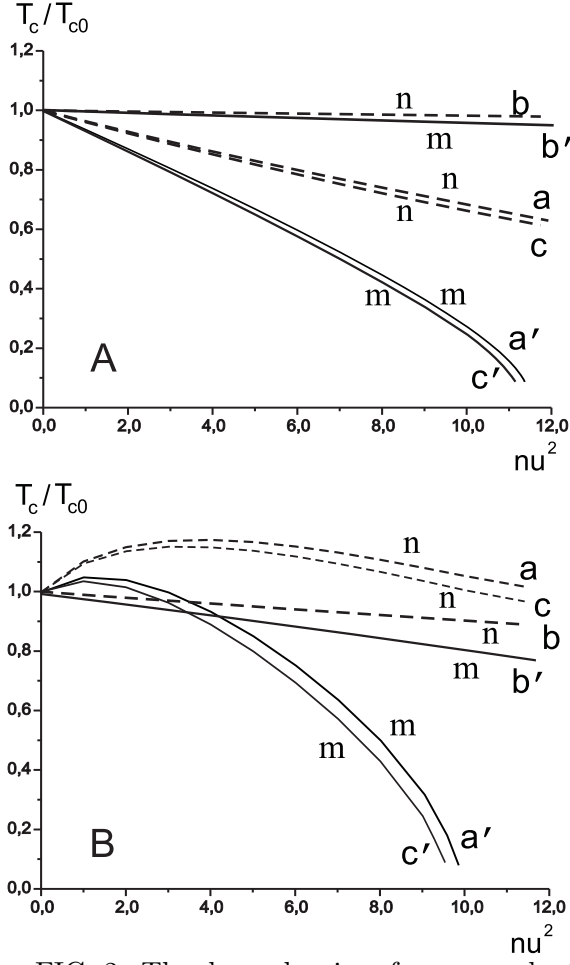}} \caption{The dependencies of superconducting
transition temperature $T_c$ on parameter $nu^2$ corresponding to electron
scattering by nonmagnetic (n) and magnetic (m) impurities calculated for
A) $\varepsilon_{c0}=0$, B) $\varepsilon_{c0}=-3$ and a) $u=u^{n}_{c}$,
$u^{n}_{ac}=u^{n}_{a}=0$; b) $u=u^{n}_{a}$, $u^{n}_{ac}=0.5u^{n}_{a}$,
$u^{n}_{c}=0$; c) $u=u^{n}_{c}$, $u^{n}_{ac}=0.5u^{n}_{c}$, $u^{n}_{a}=0$;
a') $u=u^{m}_{c}$, $u^{m}_{ac}=u^{m}_{a}=0$; b') $u=u^{m}_{a}$,
$u^{m}_{ac}=0.5u^{m}_{a}$, $u^{m}_{c}=0$; c') $u=u^{m}_{c}$,
$u^{m}_{ac}=0.5u^{m}_{c}$, $u^{m}_{a}=0$.}
\end{figure}

Fig.2 presents magnetic impurity concentration dependencies of
$T_c$. For comparison the $T_c(nu^2)$ curves for the case of
nonmagnetic impurities are also shown in the figure. If the
chemical potential is in the narrow band (Fig.2A), the quickest
reduction of $T_c$ is caused by intraband scattering in it ($a'$,
$c'$ curves). As in single-band models \cite{abr-gor}, the
critical temperature vanish at a certain impurity concentration. A
comparison of $b'$ and $c'$ curves shows that unusually weak
effect on $T_c$ is produced by interband scattering from magnetic
impurities. This fact as well as that intraband scattering
scarcely affects the critical temperature can be explained by
small value of electron density of states in the energy region
near the chemical potential and by the fact that
superconductivity is due to attraction between electrons in the
narrow band. The possibility of weak dependence of $T_c$ on
magnetic impurity concentration by considering interband
scattering was discussed earlier \cite{mazin}.

If the center of the narrow band is far from the chemical
potential (Fig.2B), then both intraband scattering in the narrow
band and interband scattering results in the usual strong
reduction of $T_c$ with increasing magnetic impurities
concentration.

The diagonalization of Hamiltonian of the considered model leads
to the problem on anisotropic electron scattering by impurities in
bands with highly anisotropic order parameters which are roughly
of $s+d$ symmetry type. The symmetry of effective scattering
matrix elements resulting from the diagonalization of the model
Hamiltonian as well as that of the order parameter are determined
by properties of electron spectrum. The degree of anisotropy of
impurity potential depends on the symmetry of initial bands and
on relative value of initial intraband and interband scattering
matrix elements. The anisotropic component could also be
introduced into initial impurity potentials, but this is beyond
the scope of this article.

Thus, within the framework of the two-band model of a
superconductor with anisotropic order parameter based on symmetry
properties of crystal lattice the impurity concentration
dependence of the critical temperature was investigated. The
performed calculations suggest that the results depend on both
the location of the chemical potential relative to the narrow
band and relative value of the intraband and interband impurity
scattering matrix elements. It was found that the critical
temperature is much less sensitive to the increase in
concentration of nonmagnetic impurities than in one-band model
with $d$-wave order parameter. As in the article \cite{kusak},
the conditions for $T_c$ increasing with impurity concentration
were found. In accordance with standard theories strong
dependence of the critical temperature on magnetic impurity
concentration was found at some values of the model parameters.

This work was supported by the RFBR Grant N02-02-16925 and Landau
Scholarship Grant.



\begin{thebibliography}{99}


\bibitem{ding}
 M.R. Norman, M. Randeria, H. Ding, J.C. Campuzano,
Phys. Rev. {\bf B 52}, 615 (1995).

\bibitem{zhao}
Y. Zhao, Phys. Rev. {\bf B 64}, 024053, (2001).

\bibitem{pon}
 Ya.G. Ponomarev, Chong Soon Khi, Kim Ki Uk, M.V. Sudakova et
 al., Physica C {\bf 315}, 85 (1999).

\bibitem{suzuk}
 M. Suzuki, T. Watanabe, A. Matsuda,
Phys. Rev. Lett. {\bf 82}, 5361 (1999).

\bibitem{anis}
 P.I. Arseyev, N.K. Fedorov , B.A. Volkov,
Sol. St. Commun., {\bf 100}, 581, (1996).

\bibitem{gor-kal}
Gor'kov, L. P., and P. A. Kalugin,  Pis'ma Zh. Eksp. Teor. Fiz. 41, 208,
(1985) [JETP Lett. 41, 253 (1985)].


\bibitem{sigr}
M. Sigrist, K. Ueda, Rev. Mod. Phys., {\bf 63}, 239, (1991).

\bibitem{maki}
Y. Sun, K. Maki Phys. Rev. {\bf B 51}, 6059, (1995).

\bibitem{abr-gor}
A. A. Abrikosov and L. P. Gor'kov, Zh. Eksp. Teor. Fiz.   39, 1781 (1960)
[Sov. Phys. JETP,  12, 1243 (1961)].



\bibitem{open}
 L.A. Openov,
  Phys. Rev. {\bf B 58}, 9468 (1998).

\bibitem{radtke}
 R.J. Radtke, K. Levin, H.-B. Schuttler, M.R. Norman,
Phys. Rev. {\bf B 48}, 653, (1993).

\bibitem{dolg}
M.L. Kulic, O.V. Dolgov, Phys. Rev. {\bf B 60}, 13062, (1999).

\bibitem{nagi}
G. Haran, A.D. Nagi, Phys. Rev. {\bf B 58}, 12441, (1998).

\bibitem{haran}
G. Haran, A.D. Nagi, Phys. Rev. {\bf B 63}, 012503, (2000).


\bibitem{gork}
D.F. Agterberg, V. Barzykin, L.P. Gor'kov, Phys. Rev. {\bf B 60}, 14868,
(1999).

\bibitem{mosk}
V.A. Moskalenko, M. E. Palistrant, Zh. Eksp. Teor. Fiz.   49, 770 (1965)
[Sov. Phys. JETP,  22, 536 (1966)].


\bibitem{kusak}
T. Kusakabe, Progr. Theor. Phys., {\bf 43}, 907, (1970).

\bibitem{keller}
R. Gajic, J. Keller, M.L. Kulic, Solid State Commun. {\bf 76}, 731,
(1990).

\bibitem{comb}
 R. Combescot, X. Leyronas,
Phys. Rev.  {\bf B 54}, 4320, (1996).


\bibitem{mazin}
A.A. Golubov, I.I. Mazin, Phys. Rev. {\bf B 55}, 15146, (1997).

\bibitem{tunn}
S.O. Loiko, N.K. Fedorov, P.I. Arseyev, Zh. Eksp. Teor. Fiz.   121, 453
(2002) [Sov. Phys. JETP,  94, 387 (2002)].


\bibitem{jayar}
B. Jayaram, S.K. Agarwal, C.V. Narasimbha Rao, A.V. Narlikar, Phys. Rev.
{\bf B 38}, 2903, (1988).

\bibitem{zhaoJPh93}
Y. Zhao, H.K. Liu, G. Yang, S.X. Dou, J. Phys.: Condens. Matter {\bf  5},
3623, (1993).




\bibitem{tolp}
S.K. Tolpygo, J.-Y. Lin, M. Gurvitch, S.Y. Hou, J.M. Philips, Phys. Rev.
{\bf B 53}, 12462, (1996).

\bibitem{kriv}
R. Killian, S. Krivenko, G. Khalliulin, P. Fulde, Phys. Rev. {\bf B 59},
14432, (1999).

\bibitem{will}
G.V.M. Williams, J.L. Tallon, R. Meinhold, Phys. Rev. {\bf B 52}, 7034,
(1995).






\end{thebibliography}
 \end{document}